# Towards a Roadmap for Trustworthy Dynamic Systems-of-Systems

Rasmus Adler, Frank Elberzhager, Julien Siebert

Fraunhofer Platz 1, 67663 Kaiserslautern, Germany
{first.last}@iese.fraunhofer.de

**Abstract.** This paper gives insights into the DynaSoS project, which aims to propose a cross-domain roadmap for systems engineering research into trustworthy dynamic systems-of-systems with autonomous components. The project follows a methodology that combines interviews and workshops with experts from various domains and a literature review. In this paper, we motivate the project, discuss five application areas with their drivers and visions, and draw initial conclusions with respect to challenges that a research roadmap should address. Our goal is to share initial results with the research community about the DynaSoS project and invite them to contribute to the derivation of the research roadmap.
**Category:** Short Research Paper

## 1      Introduction

The digital transformation is making its way into all sectors, enabling the creation of digital products, processes, and business models that are radically changing markets. Most visions of the future – be it production as a service in Industry 4.0, autonomous transportation systems in smart mobility, or cyber-physical systems in digital health – are essentially made possible by the implementation of systems-of-systems (SoS) in which several stakeholders and systems from different domains with different interests have to interoperate. In addition, there is a trend towards increased (or planned) use of autonomous systems (enabled by the application of AI and big data methods such as deep learning) and increased dynamic aspects of SoS (e.g., virtual operators, on-demand). A key characteristic of these future SoS is the high dynamism and the complex interdependencies between technology, society, and the ecological environment. For instance, on-demand shared mobility implies dynamic adaptation to mobility demands and possibly dynamic pricing according to current particle pollution status or demand/offer. Another example are smart grids, which adapt to needs from the social subsystem and use dynamic pricing as a means to influence the behavior of the social subsystem and to adapt to current weather conditions. In this paper, we focus on such

**dyna**mic **S**ystems-**o**f-**S**ystems, which we investigate in the scope of the **DynaSoS** project funded by the German Federal Ministry for Education and Research.

Software and systems engineering as it has established itself in dynaSoS application domains originated from contexts in which singular, closed systems were developed, and is therefore only partially suited to address novel challenges brought on by dynaSoS. In areas where SoS thinking is traditionally more established, new dynaSoS aspects such as the increased dynamism and the greater autonomy of components bring new challenges, such as assuring that a smart grid will not cause a power blackout.

We are witnessing various domains engaging in similar approaches to the digital transformation. Indeed, different domains are proposing roadmaps and reference architectures such as the Reference Architecture Model for Industry 4.0 (RAMI 4.0), the AUTomotive Open System Architecture (AUTOSAR), or the Smart Grid Architecture Model (SGAM). From an SoS engineering perspective, this raises the question of whether there is a risk of different domains reinventing the wheel and wasting effort, and whether SoS from multiple sectors are being developed in parallel without being able to interact. Another question is whether solutions from different domains can be adapted to each other. In summary, we observe the following:

1) Several trends are leading SoS to incorporate more and more dynamic aspects and autonomy. These aspects can provide solutions to global societal problems such as sustainability or resilience.
2) State-of-the-art SoS engineering (SoSE) needs to be enhanced to address these trends, foster cross-domain fertilization with respect to the engineering of dynaSoS, and enable cross-domain dynaSoS.

Following these observations, the DynaSoS research project is developing a research roadmap for trustworthy, dynamic systems-of-systems with autonomous components. In the course of this project, a team of 38 experts from the Fraunhofer Institute for Experimental Software Engineering IESE is performing an exhaustive literature review and conducts many interviews and workshops with representatives from industry and academia from different domains in order to collect domain-specific challenges and extract common underlying research challenges. Currently, we are still in the phase of consolidating and structuring the obtained information. However, we already want to share some of our main findings.

This paper is structured as follows: Section 2 presents some aspects of the state of the art in SoSE. Section 3 shows initial key results of our domain analyses. Section 4 discusses the two observations above and provides first examples demonstrating new research challenges. Section 5 concludes our short paper.

## 2      Related Work

The methodology of the DynaSoS project relies upon both expert knowledge and scientific literature. In the past, several roadmaps and whitepapers have been published using such a methodology. For instance, Axelband et al. proposed a list of ten research themes and their corresponding challenges in the context of SoS architecture from a workshop (13 participants). Dogan et al. proposed 78 research elements grouped into twelve major themes that emerged from T-AREA-SoS activities and were prioritized by a community of experts (38 participants). Ncube et al. discuss challenges related to requirements engineering for SoS and propose a list of ten research themes gathered from online surveys and workshops (150 participants) during the activities of four research projects (T-AREA-SoS, Road2SoS, COMPASS, and DANSE). It is also worth mentioning INCOSE's Vision 2035 (and its earlier versions: Vision 2020 and Vision 2025), which also brought together challenges and presents a roadmap in which SoS aspects are central.

Note that each SoS engineering sub-domain also offers a set of agendas based upon literature reviews. For example, 32 primary studies (31 between 2005 and 2015, plus one study from 1994) were analyzed by Lana et al. in order to identify initiatives, trends, and challenges in SoS development. In the domain of architecture, several reviews have been performed over the last ten years (e.g., Dridi et al.). Santos et al. analyzed 41 studies (between 2006 and 2021) in order to extract challenges related to architecture evaluation.

Note that traditional use cases for SoS are primarily in the fields of defense, national security, military, aerospace (for historical reasons), transportation, and energy. Use cases for SoS in the fields of smart building, smart home, smart city, or healthcare are less common. The approach followed in the Dynasos project aims at targeting application domains with lower levels of maturity than the traditional ones but with promising potential: energy, mobility, farming, manufacturing, and smart cities. Hence, it is broader (in terms of application domains) than most approaches found in the literature, but focuses more on the specific aspects of dynamics, autonomy, and trustworthiness.

## 3    DynaSoS Application Domains and Drivers

In this section, we present initial major drivers for dynaSoS in different fields of application that we have analyzed so far in the context of our DynaSoS research project in order to derive domain-independent research challenges and a related research roadmap.

**Smart energy:** A major driver in the energy sector is the necessary transition to green energy and the need to reduce $CO_2$ emissions. This inevitably leads to an increase of dynamism. Energy based on wind and solar leads to volatile sources. Furthermore, predicting energy consumption and controlling it via dynamic pricing introduces further dynamics to the smart grid. In this way, the smart grid (a technical part of the dynaSoS) dynamically interacts with the weather (an ecological part of the dynaSoS) and the end users (a social part of the dynaSoS).

**Smart mobility:** A major driver in mobility is again the reduction of $CO_2$ emissions. However, vision zero includes not only zero emissions, but also zero congestion and zero accidents. Electrification is a huge and reasonable trend, but concepts such as shared mobility, on-demand mobility, and multimodal mobility are indispensable for achieving zero emissions and zero congestion. Automated and connected driving based on intelligent transportation systems supports these concepts and can contribute to safety. The aforementioned concepts inevitably introduce dynamics into the technical part of the mobility SoS because the SoS has to adapt to the social demands. Furthermore, the technical part of the mobility dynaSoS will adapt to the current particle pollution level, which in turn depends on the mobility demands of the social system. As in the energy domain, dynamic pricing can be used to influence the behavior of mobility service users. Moreover, dynamic pricing in the energy dynaSoS may even affect dynamic pricing in the mobility dynaSoS.

**Smart farming and agri-food sectors:** The smart farming and agri-food sectors are strongly driven by sustainability goals and related regulations, like the European Green Deal, or the European Farm2Fork strategy. Swarms of partially autonomous systems such as field robots or drones can contribute to the achievement of sustainability goals in farming. First, this allows soil compaction to be reduced. Second, it allows adjusting the time spent on treating a plant and thus enables individual treatments that minimize the usage of fertilizer, fungicides, and other inputs. Third, smaller and more diverse (polycultural) fields become possible. These machines collaborate not only with each other but are connected to a farm management information system (FMIS) in order to provide data about the plants, soil, and so on. The FMIS, in turn, can be combined with an agricultural data space that interacts with various stakeholders, including farmers for decision-making and machine manufacturers for improving their machines and dynamically parameterizing them. Last but not least, it supports transparency of the food chain and provides sustainability information to supervisory authorities and end-users. The provided information can be used to dynamically adapt the overall agricultural dynaSoS.

**Smart manufacturing:** Major drivers in smart manufacturing are sustainability (cf. Industry 5.0), sovereignty, and resilience of the supply chain. A general vision for achieving this resilience is to transition from a static supply chain to a flexible supply network. This requires opening up the automation pyramid and engineering factories that can be adapted according to the needs of the supply network. This dynamic adaptation includes not only the production plant itself, but also the related flow of material. In order to achieve this flexible material flow, autonomous forklifts, autonomous mobile robots, and drones collaborate and interact with management systems within the factory. There are various triggers for adaptation at various levels in the overall manufacturing dynaSoS. The Covid-19 pandemic and the current war in Ukraine are the most recent examples of adaptation triggers with a strong impact. The bullwhip or whiplash effect shows that the dynamism of the adaptation is of crucial importance in a manufacturing dynaSoS.

**Smart city and smart (rural) regions:** Major drivers for smart city or smart rural areas are resilience, sustainability, and demographics. Cities are areas where multiple SoS need to interact. Mobility, energy, climate, and water management are only a few examples of SoS that cooperate at the city level. In addition, there is the interaction with the urban environment and its biodiversity (e.g., parks, urban gardens) and also with the social systems (e.g., culture, politics, etc.). The constraints imposed by city infrastructures (e.g., the fact that technical subsystems and infrastructure are often fixed and planned for the long term) make them a particularly interesting candidate for the application of dynaSoS.

## 4      Discussion on DynaSoS Engineering Challenges

In the previous section, we exemplified the dynamism of dynaSoS in different domains and the need to achieve objectives of crucial importance. In the following, we will present which challenges industry faces when engineering dynaSoS. Furthermore, we will analyze these challenges with respect to their potential for cross-domain fertilization, and the need to enhance state-of-the-art SoSE with respect to the characteristics of the exemplified dynaSoS.

A first challenge concerns the interoperability between systems. V2X in mobility, agrirouter in agriculture, or the asset administration shell in smart manufacturing require many stakeholders to agree on interoperability solutions. Currently, there are often only proprietary APIs and data spaces exist only at a conceptual level, like the agricultural data space. Data governance is under development and needs to deal with regulations such as the GDPR. From the interviews and our own experiences, we see this challenge rather with finding consensus than as a fundamentally technical challenge. Nevertheless, this topic has high potential for cross-domain fertilization. For instance, some concepts such as smart manufacturing (e.g., BaSys 4.0) could already be transferred to the mobility domain in the context of the project CATENA-X. We also see great potential of these solutions getting transferred to the agricultural domain.

A second challenge is related to the inappropriate structures and incentives of the involved parties that build, operate, and use dynaSoS. For instance, it is not clear which stakeholders orchestrate and shape the digital ecosystem to offer on-demand, shared, multi-modal mobility-as-a-service solutions. As long as there is no party that takes on the orchestrating role, no dynaSoS will be engineered. It might somehow emerge some time when different parties offer services that build upon each other, but this differs from engineering with respect to objectives like the sustainability development goals. For instance, a dynaSoS that is not properly engineered may result in increased mobility due to cost reduction and thus in an increase of $CO_2$ emissions. It is currently unclear who sets the main objectives and assures their achievement when shaping a digital ecosystem.

A third challenge is related to the necessary complexity of the dynaSoS to support the achievement of the sustainability development goals (SDG). SDGs are diverse and typically one choice towards behavioral change by means of regulations or implementing autonomous behavior of a technical system impacts several SDGs at once. For instance, the sustainability triangle visualizes the complex interdependencies between social, environmental, and economic aspects that need to be considered when engineering solutions that promote SGDs. It is challenging to assure that a dynaSoS within this triangle will steer the whole system towards SDGs. Who ought to be behind the steering wheel? How do we assure that these entities will actually steer the whole system in the right direction? Answering such questions requires the involvement of many research disciplines including ethics, regulatory science, economics, sociology, and computer science. Systems engineering is a transdisciplinary and integrative approach, but dynaSoS demands the involvement of disciplines that have not been considered in traditional SoSE from military or aerospace.

A fourth challenge concerns the assurance of critical properties. Engineering a smart grid that satisfies primary sustainability objectives is one thing, but assuring the absence of blackouts is another. A key characteristic of dynamism is a more efficient use of resources, but this optimization comes at the risk of not having enough back-up resources to deal with critical unforeseen events or to break chain reactions.

## 5  Summary and Conclusion

The aim of the DynaSoS project is to collect, consolidate, and present research issues and potential directions in engineering systems-of-systems. We have begun analyzing the literature and are conducting interviews with experts from different domains. We presented initial results in this short paper, focusing on key drivers of different domains and on the four initial challenges we identified so far. From our point of view, we see dynaSoS as indispensable for the future, while acknowledging that such systems readily exist in certain domains. Furthermore, when engineering such SoS, approaches, techniques, and methods have to be enhanced, or existing ones extended in order to address new challenges. We only highlighted some needs for enhancements in this short paper. Our next steps will be to continue the analyses and consolidate the gathered knowledge. We invite SoSE experts to contribute their insights to this ongoing process.

## 6  Acknowledgments

The research described in this paper was performed in the DynaSoS project (grant no. 01|S21104) of the German Federal Ministry of Education and Research (BMBF). We thank Sonnhild Namingha for proofreading.

## References

Axelband, E.; Baehren, T.; Dorenbos, D.; Madni, A.; Robitaille, P.; Valerdi, R. et al. (2007): A research agenda for systems of systems architecting. INCOSE 2007

Dogan, H.; Ncube, C.; Lim, S. L.; Henshaw, M.; Siemieniuch, C.; Sinclair, M. et al. (2013): Economic and societal significance of the systems of systems research agenda, 2013 IEEE International Conference on Systems, Man, and Cybernetics, SMC 2013

Dridi, C. E.; Benzadri, Z.; Belala, F. (2020): System of Systems Modelling: Recent work Review and a Path Forward. In ICAASE 2020 - Proceedings, 4th International Conference on Advanced Aspects of Software Engineering

Lana, C. A.; Souza, N. M.; Delamaro, M. E.; Nakagawa, E. Y.; Oquendo, F.; Maldonado, J. C. (2017): Systems-of-systems development: Initiatives, trends, and challenges. In Proceedings of the 2016 42nd Latin American Computing Conference, CLEI 2016.

Ncube, C.; Lim, S. L.; Amyot D., Maalej W., Ruhe G. (2018): On systems of systems engineering: A requirements engineering perspective and research agenda. In Proceedings - 2018 IEEE 26th International Requirements Engineering Conference, RE 2018.

Santos, Daniel S.; Oliveira, Brauner R. N.; Kazman, Rick; Nakagawa, Elisa Y. (2022): Evaluation of Systems-of-Systems Software Architectures: State of the Art and Future Perspectives. In ACM Comput. Surv.